\documentclass[preprint,12pt]{elsarticle}

\pdfoutput=1
\usepackage{graphicx}
\usepackage{dcolumn}
\usepackage{bm}
\usepackage{url}
\usepackage{amssymb}
 \expandafter\let\csname equation*\endcsname\relax 
  \expandafter\let\csname endequation*\endcsname\relax 
\usepackage{amsmath}
\usepackage[colorlinks=true, a4paper=true, pdfstartview=FitV,
linkcolor=blue, citecolor=blue, urlcolor=blue]{hyperref}

\include{epsf} 

\newcommand{\N}{{\rm I\! N}}
\newcommand{\R}{{\rm I\! R}}

\newtheorem{Def}{Definition}
\newtheorem{Proposition}[Def]{Proposition}

\newtheorem{Definition}[Def]{Definition}

\journal{Probabilistic Engineering Mechanics}

\begin{document}

\begin{frontmatter}

\title{Reliability measures for indexed semi-Markov chains
applied to wind energy production}

\author{Guglielmo D'Amico}  
\address{Dipartimento di Farmacia, 
Universit\`a `G. D'Annunzio' di Chieti-Pescara,  66013 Chieti, Italy}
\author{Filippo Petroni}
\address{Dipartimento di Scienze Economiche ed Aziendali,
Universit\`a degli studi di Cagliari, 09123 Cagliari, Italy}
\author{Flavio Prattico}
\address{Dipartimento di Ingegneria Industriale e dell'Informazione e di Economia, Universit\`a degli studi dell'Aquila, 67100
L'Aquila, Italy}
\bigskip

\begin{abstract}
The computation of the dependability measures is a crucial point in the planning and development of a wind farm.
In this paper we address the issue of energy production by wind turbine by using an indexed semi-Markov chain as
a model of wind speed. We present the mathematical model, we describe the data and technical characteristics of a commercial
wind turbine (Aircon HAWT-10kW). We show how to compute some of the main dependability measures such as reliability,
availability and maintainability functions. We compare the results of the model with real energy production obtained from data
available in the Lastem station (Italy) and sampled every 10 minutes. 
\end{abstract}

\date{\today}

\end{frontmatter}


\section{Introduction}
Wind is one of the most important renewable energy sources. Wind energy is produced by converting the kinetic energy of wind
into electrical energy by means of a generator. For this reason it is important to dispose of an efficient stochastic model for
wind speed changes.
In a recent paper \cite{dami13b} the authors showed that an indexed semi-Markov chain (ISMC) reproduces almost exactly the statistical features of wind speed. In particular, it was shown that density and autocorrelation functions of a time series of real wind speed and those obtained by the ISMC model  through Monte Carlo simulation were almost indistinguishable.

In this work we use the ISMC model to compute dependability measures as availability, reliability and maintainability functions
for the index semi-Markov process. These indicators give important information on the feasibility of
the investment in a wind farm by giving the possibility to quantify the uncertainty in the wind energy production.

Another important aspect is related to the location of the wind farm. In fact, today many wind farm are built offshore
for different reasons: the wind speed is more powerful and constant due to the absence of obstacles, the visual, environmental
and acoustic impact is cut down. The maintenance cost, instead, is higher than the onshore wind farm. A good stochastic model
can help the planning of preventive maintenance suggesting when it is suitable to do the maintenance operation. 

The results presented here are new and generalize some of the results obtained for semi-Markov chain(see \cite{barb04,Bla04,limn03,dami13a,dami09,dami11,dami11b}). The model generalizes also Markov chains and renewal models.\\

We apply our model to a real case of energy production. For this reason we choose a commercial wind turbine, a 10 kW
Aricon HAWT assumed to be installed at the station of L.S.I -Lastem which is situated in Italy.\\

The paper is organized as follows. Section 2 presents some definitions and notation on the indexed semi-Markov
chain. Section 3 describes the database and the technical characteristics of the commercial wind turbine.
Section 4 shows the way in which it is possible to compute the dependability measures via kernel transformations and the value
computed on the real data and on the synthetic data are compared. In the last section some concluding remarks and possible
extensions are presented.

\section{The indexed semi-Markov process}

In this section, we give a short review of a particular indexed semi-Markov model advanced in \cite{dami13b} as a novel
synthetic time series generation method for wind speed data which is able to capture the persistence in the wind speed data and we give new probabilistic result about the transition probability function.\\
\indent Let us consider the stochastic process
\begin{equation}
J_{-(m+1)},J_{-m},J_{-(m-1)},...,J_{-1},J_{0},J_{1},...
\end{equation}
with a finite state space $E=\{1,2,...,S\}$. In our framework the random variable $J_{n}$ describes the wind speed at the $n$-th
transition, that is at the $n$-th change of speed value and, the state space $E$, describes the discretized wind speeds, see Section 3 for a specific choice of $E$.\\
\indent Let us also consider the stochastic process
\begin{equation}
T_{-(m+1)},T_{-m},T_{-(m-1)},...,T_{-1},T_{0},T_{1},...
\end{equation}
with values in $\N$. The random variable $T_{n}$ describes the time in which the $n$-th change of value of the wind speed
occurs. We denote the stochastic process $\{X_{n}\}_{n\in \N}$ of the sojourn times in wind speed state $J_{n-1}$ before the
$n$th jump. Thus we have for all $n\in \N$ $X_{n}=T_{n+1}-T_{n}$.\\
\indent Let us consider another stochastic process
\begin{equation}
V_{-(m+1)},V_{-m},V_{-(m-1)},...,V_{-1},V_{0},V_{1},...
\end{equation}
with values in $\R$. The random variable $V_{n}$ describes the value of the index process at the $n$th transition:
\begin{equation}
\label{funcrela}
V_{n}^{m}=\frac{1}{T_{n}-T_{n-(m+1)}}\sum_{k=0}^{m}\sum_{s=1}^{X_{n-1-k}}f(J_{n-1-k},s),
\end{equation}
where $f:E\times \N \rightarrow \R$ is a generic function and $V_{-(m+1)}^{m},...,V_{0}^{m}$ are known and non-random.\\
\indent The function $f$ depends on the state of the system $J_{n-1-k}$ and on the time $s$.\\
\indent The process $V_{n}^{m}$ can be interpreted as a moving average of the accumulated reward process with the function $f$
as a measure of the permanence reward per unit time.\\
\indent It should be noted that the order of the moving average is on the number of transitions $m+1$ which is fixed. Anyway,
the moving average is executed on time windows of variable length $T_{n}-T_{n-(m+1)}=\sum_{k=n-(m+1)}^{n}X_{k}$ because each
transition has a random sojourn time $X_{k}$ of permanence in state $J_{k-1}$ before the next jump.\\
\indent The indexed model is fully specified once the dependence structure between the variables is assumed. Toward this end we
adopt the following assumption:
\begin{equation}
\label{kernel}
\begin{aligned}
& \mathbb{P}[J_{n+1}=j,\: T_{n+1}-T_{n}\leq t |\sigma(J_{h},T_{h}),\, h=-m,...,0,...,n, J_{n}=i, V_{n}^{m}=v]\\
& =\mathbb{P}[J_{n+1}=j,\: T_{n+1}-T_{n}\leq t |J_{n}=i, V_{n}^{m}=v]=:Q_{ij}^{m}(v;t),
\end{aligned}
\end{equation}
\noindent where $\sigma(J_{h},T_{h}),\, h\leq n$ represents the set of past values, of the processes $(J,T)$.
The matrix of functions ${\bf Q}^{m}(v;t)=(Q_{ij}^{m}(v;t))_{i,j\in E}$ is called $indexed$ $semi$-$Markov$ $kernel$.\\
\indent The joint process $(J_{n},T_{n})$, which is embedded in the indexed semi-Markov kernel, depends on the moving average
process $V_{n}^{m}$, the latter acts as a stochastic index. Moreover, the index process $V_{n}^{m}$ depends on $(J_{n},T_{n})$
through the functional relationship $(\ref{funcrela})$.\\
\indent Observe that if 
\begin{equation}
\mathbb{P}[J_{n+1}=j,\: T_{n+1}-T_{n}\leq t |J_{n}=i, V_{n}^{\lambda}=v]=\mathbb{P}[J_{n+1}=j,\: T_{n+1}-T_{n}\leq t |J_{n}=i],
\end{equation}
\noindent for all values $v\in \R$ of the index process, then the indexed semi-Markov kernel degenerates in an ordinary semi-Markov kernel and the model becomes equivalent to classical semi-Markov chain models as presented for example in \cite{jans06,barb08,dami12c,ciardo90,dami09b,csenk95}. The dependence of the process $(J_{n},T_{n})$ on the new variable $V_{n}^{m}$ is introduced in order to capture the effect of the past transitions on the future ones for those processes which are strongly autocorrelated.\\
\indent One of the main problem is the proposal of a particular choice of the index process which is useful in describing the wind speed process. To this end we need to choose a specific form of the function $f$.\\
The choice is motivated by some physical reasons and by model simplicity. 

Let us briefly remind that wind speed data are long range positively autocorrelated. This implies that there are periods of high and low speed. Motivated by the empirical facts in \cite{dami13b} we supposed that also the transition probabilities from current wind speed $J_{n}$ to the next one $J_{n+1}$ depends on whether the wind is, on average, in a high speed period or in a low one. We then fixed the function $f$ to be the wind speed itself, i.e. 
\begin{equation}\label{f}
f(J_{n-1-k},s)=J_{n-1-k}, 
\end{equation}
for all $s\in \N$. Consequently, substituting (\ref{f}) in equation (\ref{funcrela}) and considering that $J_{n-1-k} $ is constant in $s$ we obtain
\begin{equation}
\label{indicespecifico}
V_{n}^{m}= \frac{1}{T_{n}-T_{n-(m+1)}}\sum_{k=0}^{m}J_{n-1-k}\cdot X_{n-1-k} =\sum_{k=0}^{m} J_{n-1-k}\cdot
\bigg(\frac{T_{n-k}-T_{n-1-k}}{T_{n}-T_{n-(m+1)}}\bigg).
\end{equation}
\indent In this simple case the index process $(V_{n}^{m})$ expresses a moving average of order $m+1$ executed on the series of
the wind speed values $(J_{n-1-k})$ with weights given by the fractions of sojourn times in that wind speed
$(T_{n-k}-T_{n-1-k})$, with respect to the interval time on which the average is executed $(T_{n}-T_{n-(m+1)})$.\\
\indent The consequence of the choice (\ref{f}) is that the assumption (\ref{kernel}) now states that:
\begin{equation}
\label{kernelII}
Q_{ij}^{m}(v;t)=\mathbb{P}[J_{n+1}=j,\: T_{n+1}-T_{n}\leq t |J_{n}=i, V_{n}^{m}=\sum_{k=0}^{m}
J_{n-1-k}\bigg(\frac{T_{n-k}-T_{n-1-k}}{T_{n}-T_{n-(m+1)}}\bigg)=v].
\end{equation}
\indent Relation $(\ref{kernelII})$ asserts that the knowledge of the last wind speed value $(J_{n}=i)$ and of the weighted
moving average $(V_{n}^{m}=v)$ of order $m+1$ of past wind speeds suffices to give the conditional distribution of the couple
$J_{n+1}$, $T_{n+1}-T_{n}$ whatever the values of the past variables might be. Essentially we consider that the average of the past $m$ speeds contains most of the information needed to establish the probability of the next wind speed transition. We will show later that, with this assumption, the model is able to capture the temporal dependence structure of real data.\\
\indent We introduce now auxiliary probabilities which are helpful in the sequel of the analysis. Denote by
$$
p_{ij}^{m}(v):= \mathbb{P}[J_{n+1}=j|J_{n}=i,V_{n}^{m}=v].
$$ 
\indent They represent the transition probabilities of the embedded indexed Markov chain. More precisely $p_{ij}^{m}(v)$ denotes
the probability that the next transition is into wind speed $j$ given that at current time the wind speed process entered state $i$ and the index process had value $v$. It is simple to realize that
\begin{equation}
p_{ij}^{m}(v)=\lim_{t\rightarrow \infty}Q_{ij}^{m}(v;t).
\end{equation}
\indent Let $H_{i}^{m}(v;\cdot)$ be the sojourn time cumulative distribution in wind speed state $i\in E$:
\begin{equation}
\label{acca}
H_{i}^{m}(v;t):= \mathbb{P}[ T_{n+1}-T_{n} \leq t |  J_n=i,\, V_{n}^{m}=v ]= \sum_{j\in E}Q_{ij}^{m}(v;t).
\end{equation}
\indent It expresses the probability to change the actual wind speed $i$ in a time less or equal to $t$ given the indexed process has value $v$.\\
\indent It is useful to consider also the conditional waiting time distribution function $G$ which expresses the following probability:
\begin{equation}
\label{G}
G_{ij}^{m}(v;t):=\mathbb{P}[T_{n+1}-T_{n}\leq t \mid J_{n}=i, J_{n+1}=j,V_{n}^{m}=v].
\end{equation}
\indent It is simple to establish that
\begin{eqnarray}
&&G_{ij}^{m}(v;t)=\left\{
                \begin{array}{cl}
                       \ \frac{Q_{ij}^{m}(v;t)}{p_{ij}^{m}(v)}  &\mbox{if $p_{ij}^{m}(v)\neq 0$}\\
                         1  &\mbox{if $p_{ij}^{m}(v)=0$}.\\
                   \end{array}
             \right.
\end{eqnarray}
\indent To describe the behavior of our model at whatever time $t\in \N$ we need to define additional stochastic processes.\\
\indent Given the three-dimensional process $\{J_{n}, T_{n}, V_{n}^{m}\}$ and the indexed semi-Markov kernel ${\bf Q}^{m}(v;t)$,
we define by
\begin{equation}
\label{stocproc}
\begin{aligned}
& N(t)=\sup\{n\in \mathbb{N}: T_{n}\leq t\};\\
& Z(t)=J_{N(t)};\\
& V^{m}(t)=\frac{1}{t-T_{(N(t)-\theta)-m}}\sum_{k=0}^{m}J_{(N(t)-\theta)-k}\cdot (t\wedge
T_{(N(t)-\theta)+1-k}-T_{(N(t)-\theta)-k})
\end{aligned}
\end{equation}
where $T_{N(t)}\leq t < T_{N(t)+1}$ and $\theta =1_{\{t=T_{N(t)}\}}$.\\
\indent The stochastic processes defined in $(\ref{stocproc})$ represent the number of transitions up to time $t$, the state of the system (wind speed) at time $t$ and the value of the index process (moving average of function of wind speed) up to $t$, respectively. We refer to $Z(t)$ as an indexed semi-Markov chain.\\
\indent The process $V^{m}(t)$ extends the process $V_{n}^{m}$ because the time $t$ can be a transition or a waiting time. It is simple to realize that, $\forall m$, if $t=T_{n}$ we have that $V^{m}(t)=V_{n}^{m}$.\\
\indent Let us introduce the stochastic process
\begin{equation}
B(t)=t-T_{N(t)},
\end{equation}
which is called backward recurrence time process and denotes the time since the last transition.\\
\indent This process is very important in a semi-Markovian framework and it is well known that the transition probabilities of a semi-Markov process depend on the value of the recurrence time process, see e.g. \cite{dami09,dami11c}.\\
\indent This is due to the fact that the conditional waiting time distribution functions $(\ref{G})$ can be of any type and then, also no memoryless distributions can be used. In this case, the time since the last transition of the system (backward value) influences the system's transition probabilities; this is the so called duration effect.\\
\indent To properly assess the probabilistic behavior of the system, we need to introduce the transition probability function. To this end, it is useful to introduce the following notation: 
$$
{\bf{J}}_{-m-1}^{0}=(J_{0}, J_{-1},...,J_{-m-1} ),
$$
$$
{\bf{T}}_{-m-1}^{0}=(T_{0}, T_{-1},...,T_{-m-1}).
$$
\indent With the equality ${\bf{J}}_{-m-1}^{0}={\bf{j}}_{-m-1}^{0}$ we denote the fact that  
$$
(J_{0}=j_{0}, J_{-1}=j_{1},...,J_{-m-1}=j_{-m-1}),
$$
and similarly with ${\bf{T}}_{-m-1}^{0}={\bf{t}}_{-m-1}^{0}$ we indicate that
$$
(T_{0}=t_{0}, T_{-1}=t_{1},...,T_{-m-1}=t_{-m-1}).
$$
Finally by ${\bf{(J,T)}}_{-m-1}^{0}$ we denote the couple of ordered vectors $({\bf{J}}_{-m-1}^{0}, {\bf{T}}_{-m-1}^{0})$.\\
\begin{Definition}
The transition probability function of the indexed semi-Markov chain $Z(t)$ is the function $^{b}\phi_{({\bf{i}_{-m-1}^{0}};j)}({\bf{t}_{-m-1}^{0}};u,t)$ defined by 
\begin{equation}
\label{transprob}
^{b}\phi_{({\bf{i}_{-m-1}^{0}};j)}({\bf{t}_{-m-1}^{0}};u,t):=\,^{b}\phi_{(i_{-m-1},i_{-m},...i_{0};j)}(t_{-m-1},t_{-m},...,t_{0};u,t),
\end{equation}
where, $i_{0}, i_{-1},...,i_{-m-1}\in E$ and $t_{0}, t_{-1},...,t_{-m-1}\in \N$.
\end{Definition}
\indent The transition probability function $(\ref{transprob})$ expresses the probability the ISMC occupies state $j$ at time $t$ given that at current time $t_{0}+u$ it is in state $i_{0}$ where it entered with last transition at time $t_{0}$ having previously visited the states ${\bf{i}_{-m-1}^{-1}}$ at times ${\bf{t}_{-m-1}^{-1}}$.\\
\indent We call the transition probabilities $(\ref{transprob})$ transition probabilities with initial backward times for the ISMC model. In fact, at the current time $t_{0}+u$ the backward recurrence time process assumes value:
\[
B(t_{0}+u)=t_{0}+u-T_{N(t_{0}+u)}=t_{0}+u-T_{0}=t_{0}+u-t_{0}=u.
\]
\indent If we set $u=0$ in $(\ref{transprob})$ we have the probability
\begin{equation}
\label{phinuova}
\phi_{({\bf{i}_{-m-1}^{0}};j)}({\bf{t}_{-m-1}^{0}};t):=\mathbb{P}[Z(t)=j|J_{0}=i_{0},...,J_{-m-1}=i_{-m-1},T_{0}=t_{0},...,T_{-m-1}=t_{-m-1}].
\end{equation}
\indent which denotes the probability that the ISMC occupies state $j$ at time $t$ given that at current time $t_{0}$ it is entered into state $i_{0}$ having previously visited the states ${\bf{i}_{-m-1}^{-1}}$ at times ${\bf{t}_{-m-1}^{-1}}$.\\
\indent The following result consists in a recursive formula for computing the transition function 
$^{b}\phi_{({\bf{i}_{-m-1}^{0}};j)}({\bf{t}_{-m-1}^{0}};u,t)$ of the ISMC $Z(t)$.
\begin{Proposition}
\label{prop}
The probabilities $^{b}\phi_{({\bf{i}}_{-m-1}^{0};j)}({\bf{t}}_{-m-1}^{0};u,t)$ verify the following equation:
\begin{equation}
\label{one}
\begin{aligned}
& ^{b}\phi_{({\bf{i}}_{-m-1}^{0};j)}({\bf{t}}_{-m-1}^{0};u,t)\\
& =\delta_{i_{0}j}\frac{\Bigg(1-H_{i_{0}}^{m}\Big(\sum_{k=0}^{m}i_{-k-1}\cdot \big(\frac{t_{-k}-t_{-k-1}}{t_{0}-t_{-m-1}}\big);
t-t_{0}\Big)\Bigg)}{\Bigg(1-H_{i_{0}}^{m}\Big(\sum_{k=0}^{m}i_{-k-1}\cdot \big(\frac{t_{-k}-t_{-k-1}}{t_{0}-t_{-m-1}}\big);
u\Big)\Bigg)}\\
& +\sum_{i_{1}\in E}\sum_{t_{1}=t_{0}+u+1}^{t}
\frac{q_{i_{0}i_{1}}\big(\sum_{k=0}^{m}i_{-k-1}\cdot \big(\frac{t_{-k}-t_{-k-1}}{t_{0}-t_{-m-1}}\big); t_{1}
-t_{0}\big)}{\Bigg(1-H_{i_{0}}^{m}\Big(\sum_{k=0}^{m}i_{-k-1}\cdot \big(\frac{t_{-k}-t_{-k-1}}{t_{0}-t_{-m-1}}\big);
u\Big)\Bigg)}\,
^{b}\phi_{({\bf{i}}_{-m}^{1};j)}({\bf{t}}_{-m}^{1};t-t_{1}),
\end{aligned}
\end{equation}
\noindent where $\delta_{i_{0}j}$ is the Kronecker delta and 
\begin{equation*}
q_{ir}(v; s)=Q_{ir}(v; s)-Q_{ir}(v; s -1).
\end{equation*}
\end{Proposition}
\textbf{Proof} 
First of all let us compute the value of the index process $V^{m}(t_{0})$ given the information set
$\{{\bf{(J,T)}}_{-m-1}^{0}=({\bf{i}}_{-m-1}^{0};{\bf{t}}_{-m-1}^{0})\}$. Because $t_{0}=T_{0}$ is a transition time, we have that $\theta =1$. Moreover $T_{-m-1}=t_{-m-1}$ and $\forall k$ $T_{-k}=t_{-k}<t$. Then, we have that
\begin{equation}
\label{starstar}
V^{m}(t_{0})=\sum_{k=0}^{m}\sum_{k=0}^{m}i_{-k-1} \frac{t_{-k}-t_{-k-1}}{t_{0}-t_{-m-1}}.
\end{equation} 
Now, being the events $\{T_{1}>t\}$ and $\{T_{1}\leq t\}$ disjoint, it follows that
\begin{equation}
\label{1}
\begin{aligned}
& \mathbb{P}[Z(t)=j|T_{1}>t_{0}+u, {\bf{(J,T)}}_{-m-1}^{0}={\bf{(i,t)}}_{-m-1}^{0}]\\
& =\mathbb{P}[Z(t)=j, T_{1}> t|T_{1}>t_{0}+u, {\bf{(J,T)}}_{-m-1}^{0}={\bf{(i,t)}}_{-m-1}^{0}]\\
& +\mathbb{P}[Z(t)=j, T_{1}\leq t|T_{1}>t_{0}+u, {\bf{(J,T)}}_{-m-1}^{0}={\bf{(i,t)}}_{-m-1}^{0}].
\end{aligned}
\end{equation}
\indent Observe that
\begin{equation}
\label{2}
\begin{aligned}
& \mathbb{P}[Z(t)=j, T_{1}> t|T_{1}>t_{0}+u, {\bf{(J,T)}}_{-m-1}^{0}={\bf{(i,t)}}_{-m-1}^{0}]\\
& = \mathbb{P}[T_{1}> t|T_{1}>t_{0}+u, {\bf{(J,T)}}_{-m-1}^{0}={\bf{(i,t)}}_{-m-1}^{0}]\\
& \cdot \mathbb{P}[Z(t)=j| T_{1}> t,T_{1}>t_{0}+u, {\bf{(J,T)}}_{-m-1}^{0}={\bf{(i,t)}}_{-m-1}^{0}].
\end{aligned}
\end{equation}
\indent The first factor on the right hand side of $(\ref{2})$ is
\begin{equation}
\label{3}
\begin{aligned}
& \mathbb{P}[T_{1}> t|T_{1}>t_{0}+u,{\bf{(J,T)}}_{-m-1}^{0}={\bf{(i,t)}}_{-m-1}^{0}]\\
& =\frac{1-H_{i_{0}}^{m}\Big(\sum_{k=0}^{m}i_{-k-1} \frac{t_{-k}-t_{-k-1}}{t_{0}-t_{-m-1}};t-t_{0}\Big)}{1-H_{i_{0}}^{m}\Big(\sum_{k=0}^{m}i_{-k-1} \frac{t_{-k}-t_{-k-1}}{t_{0}-t_{-m-1}};u\Big)}.
\end{aligned}
\end{equation}
\indent The second factor on the right hand side of $(\ref{2})$ is simply
\begin{equation}
\label{4}
\mathbb{P}[Z(t)=j | T_{1}> t,T_{1}>t_{0}+u, {\bf{(J,T)}}_{-m-1}^{0}={\bf{(i,t)}}_{-m-1}^{0}]=\delta_{i_{0}j},
\end{equation}
\noindent because being $T_{1}>t$ the time of next transition exceeds $t$ and, therefore, up to $t$ the process
remains in state $i_{0}$. Consequently the probability is one if and only if $j=i_{0}$ otherwise it will be equal to zero.\\
\indent Now let us consider the computation of the second addend on the right hand side of formula $(\ref{1})$:\\ 
\begin{equation}
\begin{aligned}
& \mathbb{P}[Z(t)=j, T_{1}\leq t|T_{1}>t_{0}+u,{\bf{(J,T)}}_{-m-1}^{0}={\bf{(i,t)}}_{-m-1}^{0}]\\
& =\sum_{i_{1}\in E}\sum_{t_{1}=t_{0}+u+1}^{t} \mathbb{P}[Z(t)=j, J_{1}=i_{1}, T_{1}=t_{1}|T_{1}>t_{0}+u,{\bf{(J,T)}}_{-m-1}^{0}={\bf{(i,t)}}_{-m-1}^{0}]\\
& =\sum_{i_{1}\in E}\sum_{t_{1}=t_{0}+u+1}^{t} \mathbb{P}[Z(t)=j| J_{1}=i_{1}, T_{1}=t_{1}, T_{1}>t_{0}+u,{\bf{(J,T)}}_{-m-1}^{0}={\bf{(i,t)}}_{-m-1}^{0}] \\
& \times \mathbb{P}[J_{1}=i_{1}, T_{1}=t_{1}| T_{1}>t_{0}+u,{\bf{(J,T)}}_{-m-1}^{0}={\bf{(i,t)}}_{-m-1}^{0}]\\
& =\sum_{i_{1}\in E}\sum_{t_{1}=t_{0}+u+1}^{t}\,^{b}\phi_{(\bf{i}_{-m}^{1};j)}({\bf{t}}_{-m}^{1};t-t_{1})\\
& \times \frac{\mathbb{P}[J_{1}=i_{1}, T_{1}=t_{1}| {\bf{(J,T)}}_{-m-1}^{0}={\bf{(i,t)}}_{-m-1}^{0}]}{\mathbb{P}[T_{1}>t_{0}+u| {\bf{(J,T)}}_{-m-1}^{0}={\bf{(i,t)}}_{-m-1}^{0}]}\\
\end{aligned}
\end{equation}
\begin{equation}
\label{6}
\begin{aligned}
& =\sum_{i_{1}\in E}\sum_{t_{1}=t_{0}+u+1}^{t}
\frac{q_{i_{0}i_{1}}\big(\sum_{k=0}^{m}i_{-k-1}\cdot \big(\frac{t_{-k}-t_{-k-1}}{t_{0}-t_{-m-1}}\big); t_{1}
-t_{0}\big)}{\Bigg(1-H_{i_{0}}^{m}\Big(\sum_{k=0}^{m}i_{-k-1}\cdot \big(\frac{t_{-k}-t_{-k-1}}{t_{0}-t_{-m-1}}\big);
u\Big)\Bigg)}\,
^{b}\phi_{({\bf{i}}_{-m}^{1};j)}({\bf{t}}_{-m}^{1};t-t_{1}).
\end{aligned}
\end{equation}
\begin{flushright}
$\Box $
\end{flushright}
If we set $u=0$ in equation $(\ref{onebis})$ we obtain the evolution equation for the transition probabilities $\phi_{({\bf{i}}_{-m-1}^{0};j)}({\bf{t}}_{-m-1}^{0};t)$:
\begin{equation}
\label{onebis}
\begin{aligned}
& \phi_{({\bf{i}}_{-m-1}^{0};j)}({\bf{t}}_{-m-1}^{0};t)\\
& =\delta_{i_{0}j}\Bigg(1-H_{i_{0}}^{m}\Big(\sum_{k=0}^{m}i_{-k-1}\cdot \big(\frac{t_{-k}-t_{-k-1}}{t_{0}-t_{-m-1}}\big);
t-t_{0}\Big)\Bigg)\\
& +\sum_{i_{1}\in E}\sum_{t_{1}=t_{0}+1}^{t}
q_{i_{0}i_{1}}\big(\sum_{k=0}^{m}i_{-k-1}\cdot \big(\frac{t_{-k}-t_{-k-1}}{t_{0}-t_{-m-1}}\big); t_{1}\big)
\phi_{({\bf{i}}_{-m}^{1};j)}({\bf{t}}_{-m}^{1};t-t_{1}).
\end{aligned}
\end{equation}

\indent As it is possible to see from the previous propositions, the transition probabilities are function of the last $m+1$ states and times of transition, then our ISMC model can also be considered as a type of $m+1$ order semi-Markov chain model. Anyway higher order semi-Markov models are associated with a dramatic increase in the model parameters that rarely is possible to estimate from real data. The ISMC model overcomes this problem because the influence of the past states and times is summarized efficiently with the introduction of the index process that computes a weighted average of the past states. This is particularly evident once we remark that the $^{b}\phi_{({\bf{i}}_{-m-1}^{0};j)}({\bf{t}}_{-m-1}^{0};u,t)$ are the unknown functions of formula $(\ref{transprob})$ whereas the remaining quantities are known once we dispose of the indexed semi-Markov kernel ${\bf Q}^{m}(v;t)$ which is the only quantity to be estimated from data. From formula $(\ref{kernel})$ we see that the influence of past states and times is summarized by the index process $V^{m}(t)$. Having estimated ${\bf Q}^{m}(v;t)$ it is possible, first to compute the sojourn time cumulative distribution $H_{i}^{m}(v;\cdot)$ through formula $(\ref{acca})$ and then, by solving equation $(\ref{transprob})$ the transition probability are obtained.


\section{Database and commercial wind turbine}

As in our previous works \cite{dami13b,dami13a,dami13c} we used a free database of wind speed sampled in a weather station situated in Italy at N 45$°$ 28' 14,9'' $-$ E 9$°$ 22' 19,9'' and at 107 $m$ of altitude. The station uses a combined speed-direction anemometer at 22 $m$ above the ground. It has a measurement range that goes from 0 to 60 $m/s$, a threshold of 0,38 $m/s$ and a resolution of 0,05 $m/s$. The station processes the speed every 10 minute in a time interval ranging from 25/10/2006 to 28/06/2011.
During the 10 minutes are performed 31 sampling which are then averaged in the time interval.
In this work, we use the sampled data that represents the average of the modulus of the wind speed ($m/s$) without a considered specific direction.
This database is then composed of about  230thousands wind speed measures ranging from 0 to 16 $m/s$.

To be able to model the wind speed as a semi-Markov process, the state space of wind speed has been discretized. In the example shown in this work we discretize wind speed into 8 states (see Table \ref{st}) chosen to cover all the wind speed distribution.
This choice is done by considering a trade off between accuracy of the description of the wind speed distribution and number of parameters to be estimated. An increase of the number of states better describes the process but requires a larger dataset to get reliable estimates and it could also be not necessary for the accuracy needed in forecasting future wind speeds. Note also that, in the database used in this work, there are very few cases where the wind speed exceeds $7m/s$. We stress that the discretization should be chosen according to the database to be used.
\begin{table}
\begin{center}

\begin{tabular}{|c|*{3}{c|}|}
     \hline
Sate & Wind speed range $m/s$ \\ \hline
1 & 0 to 1  \\ \hline
2 & 2 \\ \hline
3 & 3 \\ \hline
4 & 4 \\ \hline
5 & 5 \\ \hline
6 & 6 \\ \hline
7 & 7 \\ \hline
8 & $>$7 \\ \hline
\end{tabular} 
\caption{Wind speed discretization}
\label{st} 
\end{center}
\end{table}

\begin{figure}
\centering
\includegraphics[height=7cm]{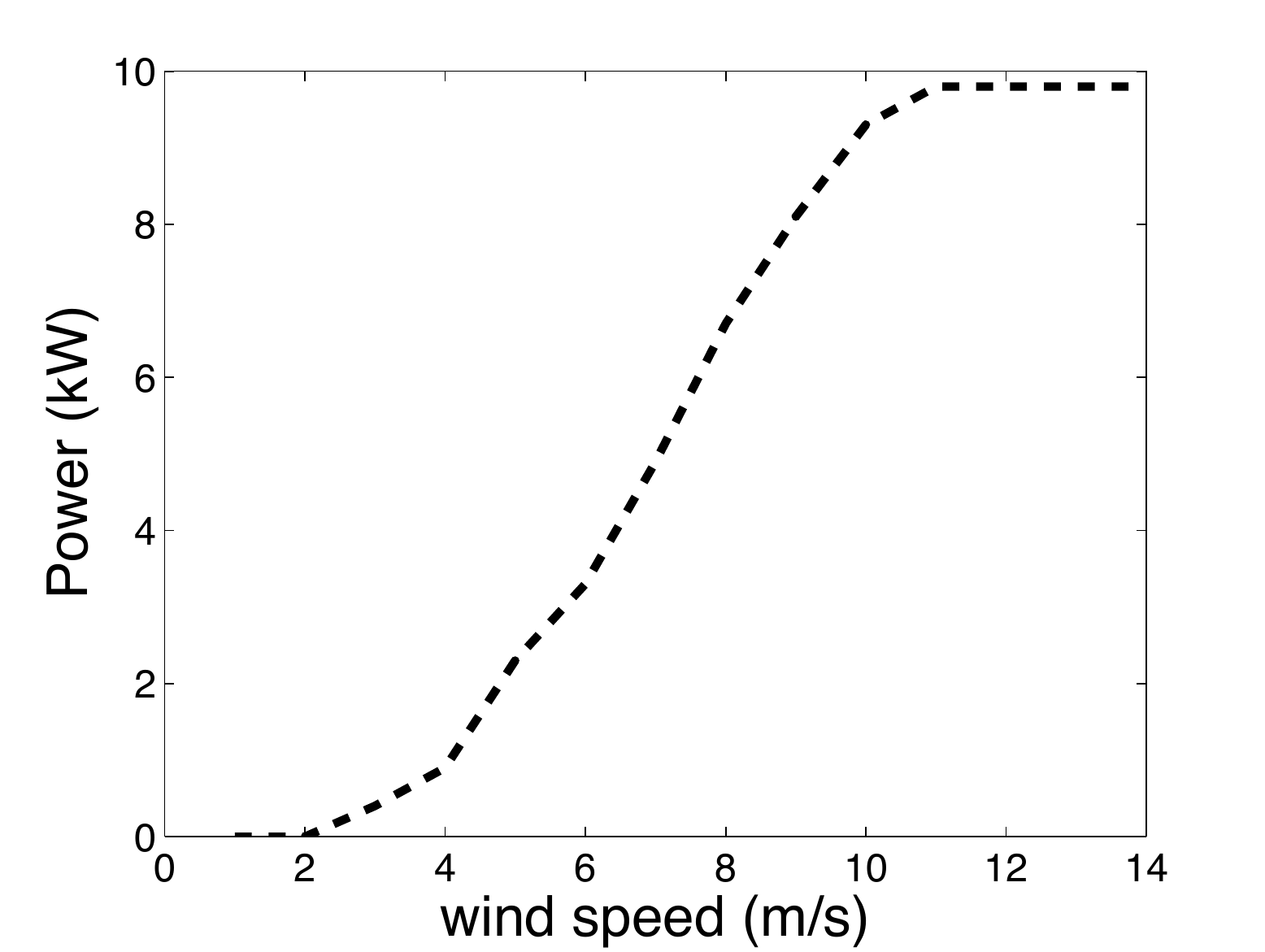}
\caption{Power curve of the 10 kW Aricon HAWT wind turbine.}\label{power}
\end{figure}

We apply our model to a real case of energy production. For this reason we choose a commercial wind turbine, a 10 kW Aircon HAWT with a power curve given in Figure \ref{power}. The power curve of a wind turbine represents how much energy it produces as a function of the wind speed. In this case, see Figure \ref{power}, there is a cut in speed at 2 $m/s$, instead the wind turbine produces energy almost linearly from 3 to 10 $m/s$, then, with increasing wind speed the energy production remains constant until the cut out speed at 32 $m/s$, in which the wind turbine is stopped for structural reason. Then the power curve acts as a filter for the wind speed. In the database used for our analysis the wind speed does never exceed 16 $m/s$ and it is seldom over 8 $m/s$, this is why the discretization is performed according to Table \ref{st} and the analysis never reached the cut out
speed.

Through this power curve we can know how much energy is produced as a function of the wind speed at a given time.

\section{Reliability theory for the ISMC model of wind speed}

In this section we define and compute reliability measures for the ISMC model.\\
Let $E$ be partitioned into sets $U$ and $D$, so that:
$$
E = U \cup D,\,\,\,\emptyset  = U \cap D,\,\,\,U \ne \emptyset ,\,\,\,U \ne E.
$$
The subset $U$ contains all "Up" states in which the system is working and subset $D$
all "Down" states in which the system is not working well or has failed. In the wind speed model the Up states are those for which the wind speed is sufficiently high to allow the production of energy or not excessive high such that the turbine should be turned off.\\
\indent In the following we present both the typical indicators used in reliability theory and also their application.\\ 
\indent The three indicators that we evaluate are the availability, reliability and maintainability functions, they were extensively studied  among others by \cite{barl75,lisn03} and related to semi-Markov models by \cite{barb04,Bla04,dami09b,dami13a,pere02,limn99}.

(i) {\it the point wise availability function $^{b}A$ with backward time} giving the probability that the system is working on at time $t$ whatever happens on $(0,t]$.\\
In our model the availability is defined as follows: 
\begin{equation}
\label{ava}
\begin{aligned}
& ^{b}A_{{\bf{i}}_{-m-1}^{0}}({\bf{t}}_{-m-1}^{0};u,t)\\ 
& = \mathbb{P}\big[Z(t)\in U|T_{1}>t_{0}+u, {\bf{(J,T)}}_{-m-1}^{0}=({\bf{(i,t)}}_{-m-1}^{0}) \big],
\end{aligned}
\end{equation}
and gives the probability that at time $t$ the wind turbine produces energy conditional on the last $m+1$ wind speed values ${\bf{i}}_{-m-1}^{0}$ registered at the times ${\bf{t}}_{-m-1}^{0}$ and on the no occurrence of next transition up to time $t_{0}+u$.\\

(ii) {\it the reliability function $^{b}R$ with backward time}
giving the probability that the system was always working from time $t_{0}$ to time $t$:
In our model the availability is defined as follows: 
\begin{equation}
\label{rel}
\begin{aligned}
& ^{b}R_{{\bf{i}}_{-m-1}^{0}}({\bf{t}}_{-m-1}^{0};u,t)\\ 
& = \mathbb{P}\big[Z(u)\in U, \forall u \in ( t_{0},t]|T_{1}>t_{0}+u, {\bf{(J,T)}}_{-m-1}^{0}={\bf{(i,t)}}_{-m-1}^{0} \big]
\end{aligned}
\end{equation}
and gives the probability that the wind turbine will always produce energy from time $t_{0}$ up to time $t$ conditional on the last $m+1$ wind speed values ${\bf{i}}_{-m-1}^{0}$ registered at the times ${\bf{t}}_{-m-1}^{0}$ and on the no occurrence of next transition up to time $t_{0}+u$.\\

(iii) {\it the maintainability function $^{b}M$ with backward time} giving the probability that the system will leave the set $D$ within the time $t$ being in $D$ at time $t_{0}$:
\begin{equation}
\label{man}
\begin{aligned}
& ^{b}M_{{\bf{i}}_{-m-1}^{0}}({\bf{t}}_{-m-1}^{0};u,t)\\ 
& = \mathbb{P}\big[\exists u \in ( t_{0},t] : Z(u)\in U |T_{1}>t_{0}+u, {\bf{(J,T)}}_{-m-1}^{0}={\bf{(i,t)}}_{-m-1}^{0} \big]
\end{aligned}
\end{equation}
and gives the probability that the turbine will produce energy at least once from time $t_{0}$ up to time $t$ conditional on the last $m+1$ wind speed values ${\bf{i}}_{-m-1}^{0}$ registered at the times ${\bf{t}}_{-m-1}^{0}$ and on the no occurrence of any transition up to time $t_{0}+u$.\\

\indent These three probabilities can be computed in the following way if the process is an indexed semi-Markov chain of kernel ${\bf Q}^{m}(v;t)$.\\

(i) {\it the point wise availability function}
$
^{b}A_{{\bf{i}}_{-m-1}^{0}}({\bf{t}}_{-m-1}^{0};u,t):
$\\
to compute these probabilities it is sufficient to use the following formula:
\begin{equation}
\label{avaleq}
^{b}A_{{\bf{i}}_{-m-1}^{0}}({\bf{t}}_{-m-1}^{0};u,t) = \sum_{j \in U} \,
^{b}\phi_{({\bf{i}}_{-m-1}^{0};j)}({\bf{t}}_{-m-1}^{0};u,t) .
\end{equation}

(ii) {\it the reliability function}
 $
^{b}R_{{\bf{i}}_{-m-1}^{0}}^{m}({\bf{t}}_{-m-1}^{0};u,t)
$

\noindent to compute these probabilities, we will now work with another cumulated kernel ${\bf \hat{Q}}^{m}(v;t)$ for which all the states of the subset $D$ are changed into absorbing states by considering the following transformation:
\begin{equation}
\label{transrelia}
\hat{p}_{i,j}^{m}(v)  = {\text{ }}\left\{ {\begin{array}{*{20}c}
   {1 \,\,\,\,\,\,{\text{if}}\,\, i\in D,\,j=i }  \\
   {0 \,\,\,\,\,\,{\text{if}}\,\, i\in D,\, j\notin D }  \\
   {p_{i,j}^{m}(v)\,\,\,\,\,\,\,{\text{otherwise.}} } \\
 \end{array} } \right.
\end{equation}
$
 ^{b}R_{{\bf{i}}_{-m-1}^{0}}({\bf{t}}_{-m-1}^{0};u,t)
$
is given by solving the evolution equation of the indexed semi-Markov chain but now with the kernel $\hat{Q}_{ij}^{m}(v;t)=
\hat{p}_{ij}^{m}(v)G_{ij}^{m}(v;t)$.\\
\indent The related formula will be:
\begin{equation}
\label{releq}
^{b}R_{({\bf{i}}_{-m-1}^{0};j)}({\bf{t}}_{-m-1}^{0};u,t) = \sum\limits_{j \in U}        	
\,^{b}\hat{\phi}_{{\bf{i}}_{-m-1}^{0}}({\bf{t}}_{-m-1}^{0};u,t)
\end{equation}
where 
$
^{b}\hat{\phi}_{({\bf{i}}_{-m-1}^{0};j)}({\bf{t}}_{-m-1}^{0};u,t)
$
are the transition probabilities of the process with all the states in 
$
D
$
 that are absorbing, i.e. with cumulated kernel ${\hat{\mathbf{Q}}}$.

(iii) {\it the maintainability function} $^{b}M_{{\bf{i}}_{-m-1}^{0}}({\bf{t}}_{-m-1}^{0};u,t) $:\\
\noindent to compute these probabilities we will now work with another cumulated kernel
${\tilde{\mathbf{Q}}}=(\tilde{Q}_{ij}^{m}(v;t))$
for which all the states of the subset $U$ are changed into absorbing states by considering the following transformation:
\begin{equation}
\label{transmain}
\tilde{p}_{ij}^{m}(v;t)  = {\text{ }}\left\{ {\begin{array}{*{20}c}
   {1 \,\,\,\,\,\,{\text{if}}\,\, i\in U,\,j=i }  \\
   {0 \,\,\,\,\,\,{\text{if}}\,\, i\in U,\,j\neq i }  \\
   {p_{ij}^{m}(v;t)\,\,\,\,\,\,\,{\text{otherwise.}} } \\
 \end{array} } \right.
\end{equation}
$
^{b}M_{{\bf{i}}_{-m-1}^{0}}({\bf{t}}_{-m-1}^{0};u,t)
$
is given by solving the evolution equation of an indexed semi-Markov chain but now with the cumulated kernel
$\tilde{Q}_{ij}^{m}(v;t)=\tilde{p}_{ij}^{m}(v)\cdot G_{ij}^{m}(v;t)$.\\
 \indent The related formula for the maintainability function will be:
\begin{equation}
\label{manteq}
^{b}M_{{\bf{i}}_{-m-1}^{0}}({\bf{t}}_{-m-1}^{0};u,t)
 = \sum\limits_{j \in U}\,^{b}\tilde{\phi}_{({\bf{i}}_{-m-1}^{0};j)}({\bf{t}}_{-m-1}^{0};u,t)
 \end{equation}
where 
$
^{b}\tilde{\phi}_{({\bf{i}}_{-m-1}^{0};j)}({\bf{t}}_{-m-1}^{0};u,t)
$
  are the transition probability of the process with all the states in 
$
U
$
 that are absorbing, i.e. with cumulated kernel ${\tilde{\mathbf{Q}}}$.\\
 
Unfortunately it is impossible to give a graphical representation of these three indicators because they depends on too many parameters ($m+1$ states ${\bf{i}}_{-m-1}^{0}$, $m+1$ times ${\bf{t}}_{-m-1}^{0}$ and one backward value $v$), then in order to show the capacity of the model to correctly reproduce the reliability measures we define unconditional reliability measures and we compute them by evaluating the discrepancies between real data and model results.\\
\indent Let us define the set
$$
\mathcal{H}_{x_{0},t_{0}}(v)=\{ ({\bf{i}}_{-m-1}^{0}, {\bf{t}}_{-m-1}^{0})\mid i_{0}=x_{0}, t_{0}=y_{0}, V^{m}(t_{0})=v\},
$$
then we define the unconditional availability function
\begin{equation}
\label{avaunc}
\begin{aligned}
& ^{b}A_{\mathcal{H}_{x_{0},t_{0}}(v)}(u,t)\\ 
& = \mathbb{P}\big[Z(t)\in U|T_{1}>t_{0}+u, {\bf{(J,T)}}_{-m-1}^{0}=({\bf{(a,s)}}_{-m-1}^{0})\in \mathcal{H}_{x_{0},t_{0}}(v) \big].
\end{aligned}
\end{equation}
The unconditional availability gives the probability that at time $t$ the wind turbine produces energy conditional on the occupancy at the current time $t_{0}$ of the state $i_{0}$ with the index process having value $v$ and the backward value equal to $u$. By using properties of the conditional probabilities it is easy to realize that
\begin{equation}
\begin{aligned}
& ^{b}A_{\mathcal{H}_{x_{0},t_{0}}(v)}(u,t)
= \sum_{({\bf{(a,s)}}_{-m-1}^{0})\in \mathcal{H}_{x_{0},t_{0}}(v)} \mathbb{P}[{\bf{(J,T)}}_{-m-1}^{0}=({\bf{(a,s)}}_{-m-1}^{0})]\\
& \times \mathbb{P}\big[Z(t)\in U|T_{1}>t_{0}+u, {\bf{(J,T)}}_{-m-1}^{0}=({\bf{(a,s)}}_{-m-1}^{0}) \big]\\
& = \sum_{({\bf{(a,s)}}_{-m-1}^{0})\in \mathcal{H}_{x_{0},t_{0}}(v)} \mu({\bf{(a,s)}}_{-m-1}^{0}) 
\,\,\,^{b}A_{{\bf{a}}_{-m-1}^{0}}({\bf{s}}_{-m-1}^{0};u,t).
\end{aligned}
\end{equation}
\noindent where $\mu({\bf{(a,s)}}_{-m-1}^{0})$ is the initial distribution of the states ${\bf{a}}_{-m-1}^{0}$ occupied at times ${\bf{s}}_{-m-1}^{0}$.\\
\indent The same definitions apply for the reliability and the maintainability functions.\\
\indent We define the unconditional reliability function
\begin{equation}
\label{avaunc}
\begin{aligned}
& ^{b}R_{\mathcal{H}_{x_{0},t_{0}}(v)}(u,t)\\ 
& = \mathbb{P}\big[Z(u)\in U, \forall u \in ( t_{0},t]|T_{1}>t_{0}+u, {\bf{(J,T)}}_{-m-1}^{0}=({\bf{(a,s)}}_{-m-1}^{0})\in \mathcal{H}_{x_{0},t_{0}}(v) \big].
\end{aligned}
\end{equation}
and it results that 
\begin{equation}
^{b}R_{\mathcal{H}_{x_{0},t_{0}}(v)}(u,t) = \sum_{({\bf{(a,s)}}_{-m-1}^{0})\in \mathcal{H}_{x_{0},t_{0}}(v)} \mu({\bf{(a,s)}}_{-m-1}^{0}) 
\,\,\,^{b}R_{{\bf{a}}_{-m-1}^{0}}({\bf{s}}_{-m-1}^{0};u,t).
\end{equation}
\indent The unconditional maintainability function is defined as follows
\begin{equation}
\label{avaunc}
\begin{aligned}
& ^{b}M_{\mathcal{H}_{x_{0},t_{0}}(v)}(u,t)\\ 
& = \mathbb{P}\big[\exists u \in ( t_{0},t] : Z(u)\in U|T_{1}>t_{0}+u, {\bf{(J,T)}}_{-m-1}^{0}=({\bf{(a,s)}}_{-m-1}^{0})\in \mathcal{H}_{x_{0},t_{0}}(v) \big].
\end{aligned}
\end{equation}
and it results that 
\begin{equation}
^{b}M_{\mathcal{H}_{x_{0},t_{0}}(v)}(u,t) 
= \sum_{({\bf{(a,s)}}_{-m-1}^{0})\in \mathcal{H}_{x_{0},t_{0}}(v)} \mu({\bf{(a,s)}}_{-m-1}^{0}) 
\,\,\,^{b}M_{{\bf{a}}_{-m-1}^{0}}({\bf{s}}_{-m-1}^{0};u,t).
\end{equation}
\indent The unconditional reliability measures can be graphically represented because if we set the current time $t_{0}=0$, then, they depend only on three parameters $(i_{0},u,v)$.  In order to verify the validity of our model, we compare the behaviour of these indicators for real and synthetic data. The indicators of the synthetic data are computed averaging over 500 different trajectories. The number of trajectories is chosen to have stable results.\\
The unconditional reliability measures are plotted  by varying the backward time \textit{u}, the index process \textit{v} and maintaining constant the initial state $i_{0}$. The numeric choice of each parameter is given only for graphical reasons, in order to show the maximum number of curves without overlaps. 
For all the figures we use the notation of indicating with a continuous line the indicators computed from real data and with a dashed line those computed from simulated data. As numeric indicator to compare the gap between the curves we compute the root mean square error (RMSE) between the indicator applied to the real data and the 500 simulated trajectories.
The mean square error is defined as follows:
\begin{equation}
\label{mse}
^{b}RMSE_{i,v}= \sqrt{\frac{1}{500}\sum_{h=1}^{500}\big(\,^{b}I_{i,v}^{real}- \,  ^{b}I_{i,v}^{h}\big)^{2}}
\end{equation}
where $^{b}I_{i,v}^{real}$ stands for the indicators estimated form real data and $^{b}I_{i,v}^{h}$ the indicators estimated from each synthetic trajectory.

\begin{figure}
\centering
\includegraphics[height=8cm]{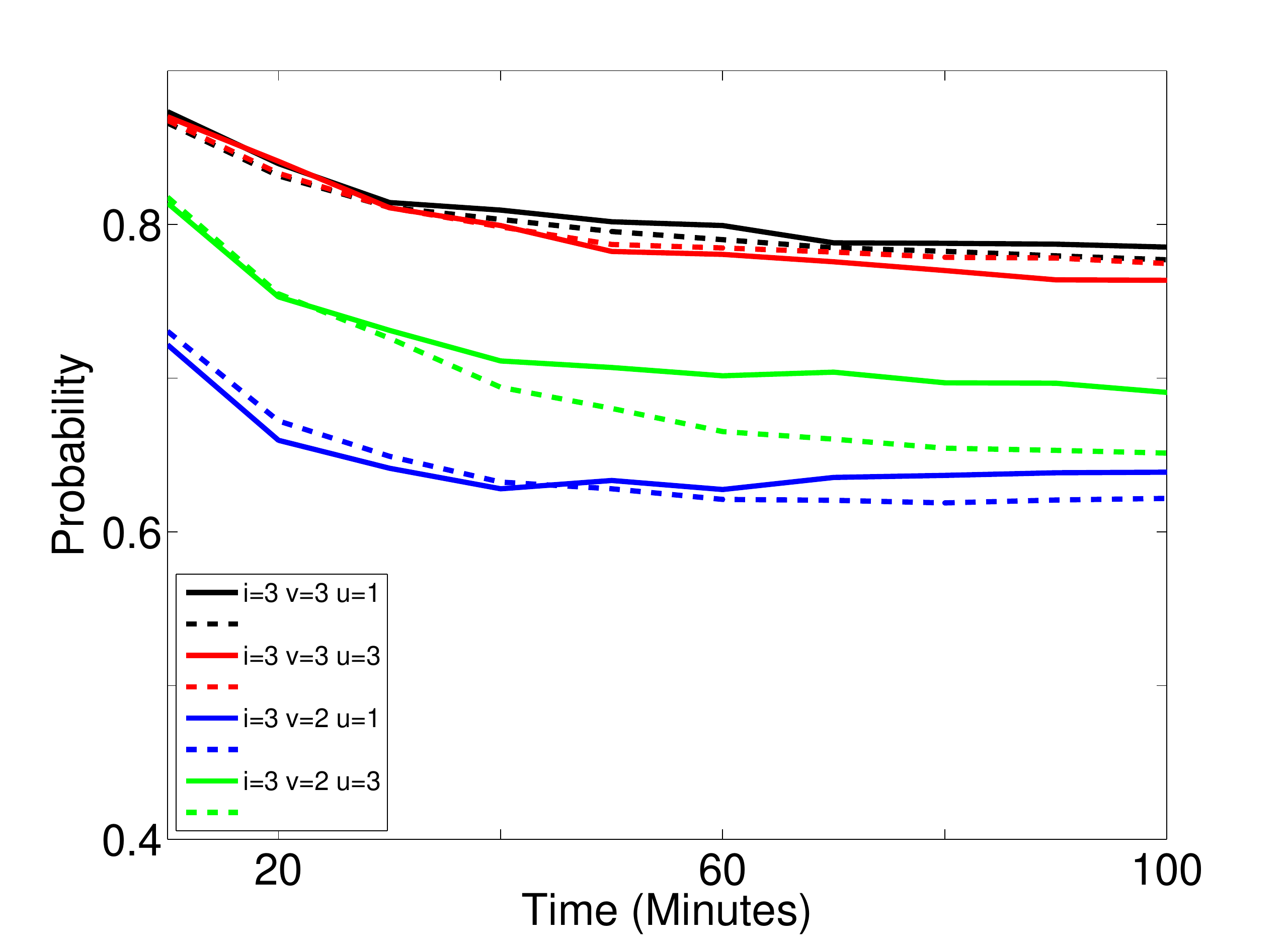}
\caption{Comparison of unconditional availability functions for real and simulated data}\label{av}
\end{figure}

\begin{table}
\begin{center}

\begin{tabular}{|c|*{3}{c|}}
     \hline

 & 10 $min$ & 50 $min$  & 100 $min$ \\ \hline
$ ^{1}RMSE_{1,3} $ & 0.0079 & 0.0072 & 0.0092  \\ \hline
$ ^{3}RMSE_{3,3} $ & 0.0044 & 0.0068 & 0.0119 \\ \hline
$ ^{1}RMSE_{3,2}$ & 0.0103 & 0.0085 & 0.0188 \\ \hline
$ ^{3}RMSE_{3,2} $ & 0.0074 & 0.0277 & 0.0405 \\ \hline

\end{tabular} 
\caption{Mean square error between the curves of the availability applied to real and synthetic data}
\label{eav} 
\end{center}
\end{table}

In Figure \ref{av} the unconditional availability functions of real and synthetic data are compared. The comparison is made
by varying the backward time \textit{u} and the index process \textit{v} and maintaining constant the starting state $i_{0}$. In Table \ref{eav} there are the RMSE values computed between real and simulated curves of Figure \ref{av}. As it is possible to note there is a slight increasing trend of the RMSE at the increasing of time. This consideration can be extended also to the unconditional reliability and maintainability functions.

\begin{figure}
\centering
\includegraphics[height=8cm]{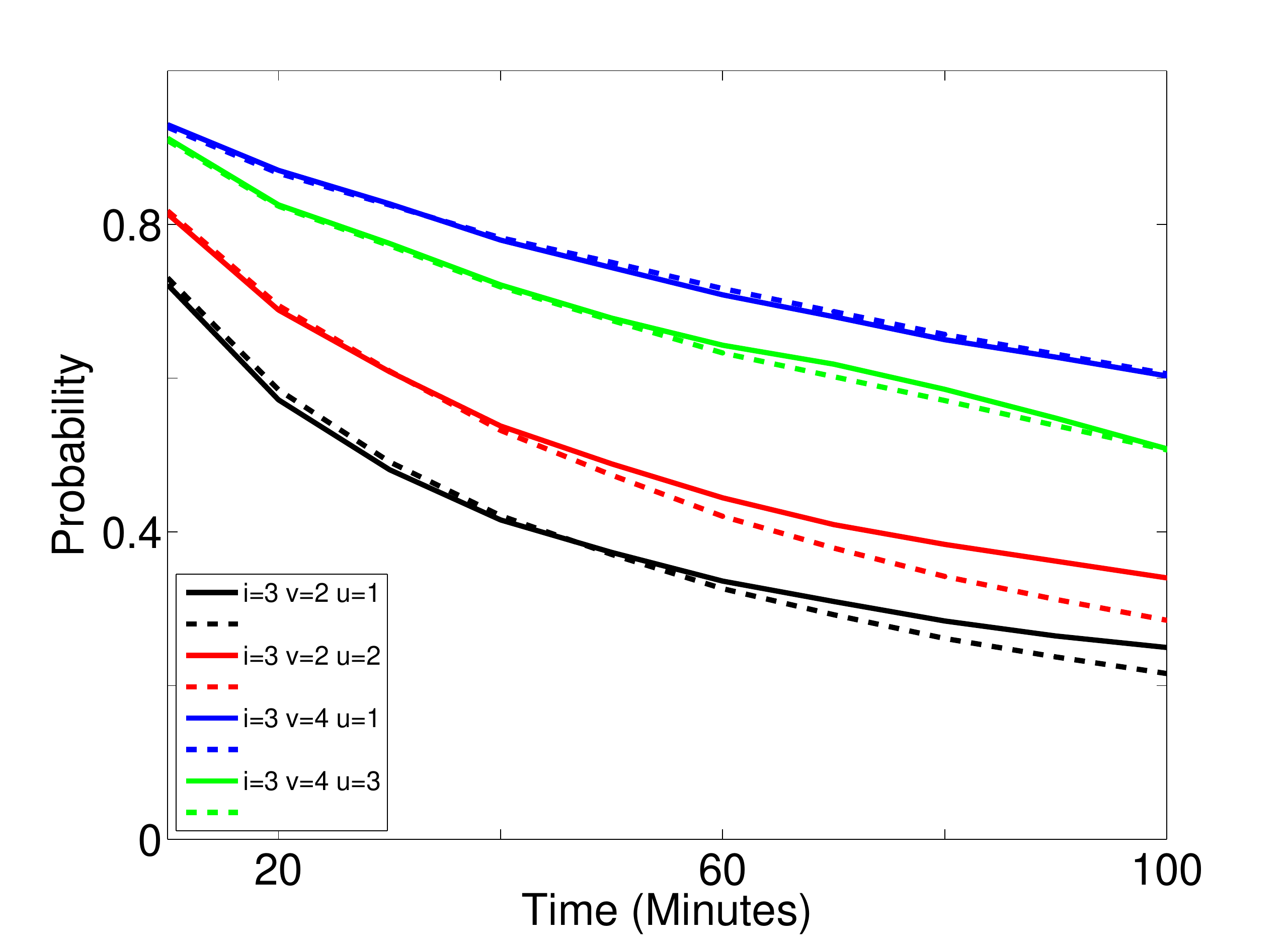}
\caption{Comparison of unconditional reliability functions for real and simulated data}\label{re}
\end{figure}

\begin{table}
\begin{center}

\begin{tabular}{|c|*{3}{c|}}
     \hline

 & 10 $min$ & 50 $min$  & 100 $min$ \\ \hline
$ ^{1}RMSE_{3,2} $ & 0.0103 & 0.0070 & 0.0344  \\ \hline
$ ^{2}RMSE_{3,2} $ & 0.0074 & 0.0170 & 0.0559 \\ \hline
$ ^{1}RMSE_{3,4}$ & 0.0065 & 0.0115 & 0.0135 \\ \hline
$ ^{3}RMSE_{3,4} $ & 0.0089 & 0.0158 & 0.0188 \\ \hline

\end{tabular} 
\caption{Mean square error between the curves of the reliability applied to real and synthetic data}
\label{ere} 
\end{center}
\end{table}
 
Figure \ref{re} shows the reliability function for real data compared with those simulated. The plotting procedure
is the same as for the previous figure.
The theoretical trend of the reliability function is confirmed, the probability decreases at the increasing of the time interval. In Table \ref{ere}, instead there are the RMSE values computed between real and simulated curves of Figure \ref{re}.
\begin{figure}
\centering
\includegraphics[height=8cm]{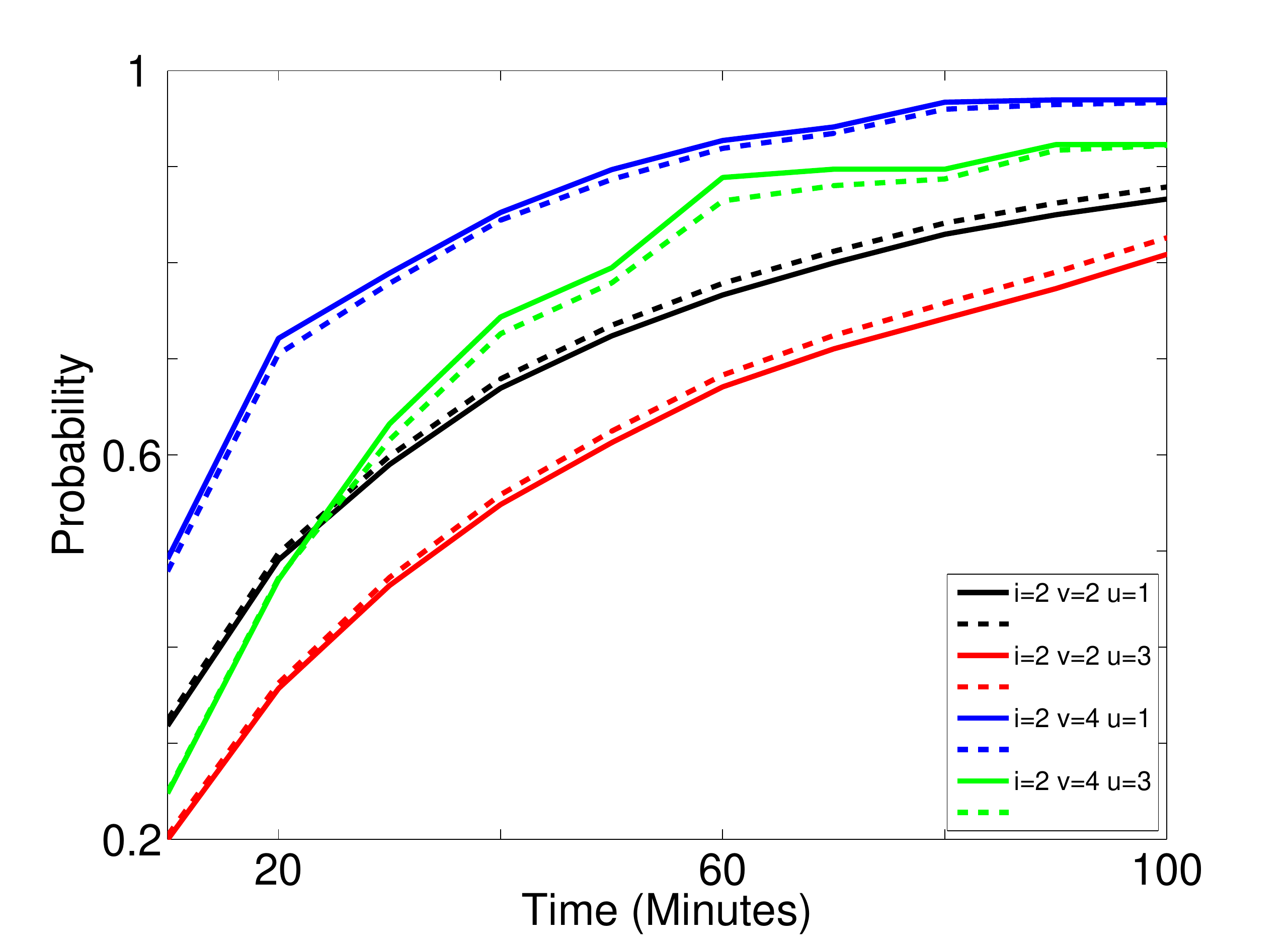}
\caption{Comparison of unconditional maintainability functions for real and simulated data}\label{ma}
\end{figure}

\begin{table}
\begin{center}

\begin{tabular}{|c|*{3}{c|}}
     \hline
 & 10 $min$ & 50 $min$  & 100 $min$ \\ \hline
$ ^{1}RMSE_{2,2} $ & 0.0093 & 0.0134 & 0.0137  \\ \hline
$ ^{3}RMSE_{2,2} $ & 0.0088 & 0.0161 & 0.0189 \\ \hline
$ ^{1}RMSE_{2,4}$ & 0.0072 & 0.0175 & 0.0087 \\ \hline
$ ^{3}RMSE_{2,4} $ & 0.0095 & 0.0377 & 0.0216 \\ \hline
\end{tabular} 
\caption{Mean square error between the curves of the maintainability applied to real and synthetic data}
\label{ema} 
\end{center}
\end{table}

The maintainability function is plotted in Figure \ref{ma}. As the previous figures, this one shows the comparison of the maintainability for real and simulated data varying the backward time and the index process and maintaining constant the starting state. Table \ref{ema} shows the RMSE values computed between real and simulated curves of Figure \ref{ma}.

It is possible to note that all the indicators plotted above (availability, reliability and maintainability)
depend strongly on the backward time and on the index process. 
In fact, all the probabilities have different values also if only the index process $v$ is changed keeping constant the backward time \textit{b} and the starting state \textit{i}. For example, from Figure \ref{re} it is possible to see that for all $s$
$$
^{b}R_{\mathcal{H}_{3,0}(4)}(1,s) > \,^{b}R_{\mathcal{H}_{3,0}(2)}(1,s) \, \forall s\in [0,100],
$$
and in particular $^{b}R_{\mathcal{H}_{3,0}(4)}(1,40)=0,77$ and $^{b}R_{\mathcal{H}_{3,0}(2)}(1,40)=0,41$. This reveals that it is important to dispose of a model that is able to distinguish between these different situations which are determined only from a different duration of permanence in the initial state $i_{0}$ and different values of the average of past $m+1$ visited states. Models based on Markov chains or classical semi-Markov chain are unable to capture this important effect that our indexed semi-Markov chain reproduces accordingly to the real data. To motivate well this statement, in Table \ref{confr} we show the percentage RMSE between all the indicators (availability, reliability and maintainability) evaluated for real and synthetic data computed through different models. Particularly we use, for this comparison, a first order Markov chain, a first order semi-Markov chain, a second order in state and duration semi-Markov chain and the ISMC model. 
\begin{table}
\begin{center}
\begin{tabular}{|l|*{4}{c|}}
     \hline
  & Markov & semi-Markov(1) & semi-Markov(2) & ISMC  \\ \hline
Availability & 65\% & 54\% & 12\% & 8\% \\ \hline
Reliability & 83\% & 69\% & 27\% & 9\% \\ \hline
Maintainability & 73\% & 65\% & 25\% & 3\% \\ \hline
\end{tabular} 
\caption{Root mean square error percentage between real and synthetic data for availability, reliability and maintainability function computed with different models}
\label{confr} 
\end{center}
\end{table}
From the Table it is clear that the best model is the ISMC model.

\section{Conclusion}
In our previous work, we presented a new stochastic model for the generation of synthetic wind speed data based on a semi-Markov approach but including a new and important random variable that able us to capture well the behaviour of the wind speed. Here we compute, for the first time, typical indicators in reliability theory for wind speed phenomenon by using the ISMC model. 

In order to check the validity of the presented model, we have compared the behaviour
of these indicators for real and simulated data. To do this, we applied our model to a real case of energy production filtering
real and simulated data by the power curve of a commercial wind turbine. The results show that the proposed model is able to
reproduce the behaviour of real data by exhibiting the dependence of the reliability indicators on the backward time and the index process.\\
\indent The indications provided by the model are of importance for assessing the suitability of a location for the wind farm
installation as well as for the planning of a preventive maintenance policy. We have also shown that the ISMC model reproduces the behavior of real data much better than a first order Markov chain, a first order semi-Markov chain and a second order in state and duration semi-Markov chain.



\begin{thebibliography}{99}


\bibitem{dami13b} G. D'Amico, F. Petroni, and F. Prattico, Wind speed modeled as an indexed semi-Markov process, {\em Environmetrics}  $\bf 24$ (2013) 367-376.

\bibitem{barb04}
V. Barbu, M. Boussemart and N. Limnios, Discrete Time Semi-Markov Model for Reliability and Survival Analysis, {\em Communications in Statistics, Theory and Methods} $\bf 33$ (2004) 2833-2868.

\bibitem{Bla04}
A. Blasi, J. Janssen, R. Manca, Numerical treatment of homogeneous and non-homogeneous semi-Markov reliability models. {\em Communications in Statistics, Theory and Methods} $\bf 33$ (2004) 697-714.

\bibitem{limn03} N. Limnios, G. Opri\c{s}an, An introduction to Semi-Markov Processes with Application to Reliability, In D.N.
Shanbhag and C.R. Rao, eds., { \em Handbook of Statistics} $\bf 21$ (2003) 515-556.

\bibitem{dami13a} G. D'Amico, F. Petroni, and F. Prattico, Reliability Measures of Second-Order Semi-Markov Chain Applied to Wind Energy Production, {\em Journal of Renewable Energy} (2013) Article ID 368940.

\bibitem{dami09}
G. D'Amico, J. Janssen,  and R. Manca, Semi-Markov Reliability Models with Recurrence Times and 
Credit Rating Applications, {\em Journal of Applied Mathematics and Decision Sciences} Article ID 625712, (2009) 17 pages.

\bibitem{dami11}
G. D'Amico, F. Petroni, A semi-Markov model with memory for price changes, {\em Journal of Statistical Mechanics} P12009 (2011).

\bibitem{dami11b} G. D'Amico, Age-usage semi-Markov models, {\em Applied Mathematica Modelling}, $\bf 35$ (2011) 4354-4366.

\bibitem{jans06} J. Janssen and R. Manca, Applied semi-Markov processes, {\em Springer}, New York, 2006.

\bibitem{barb08} V. Barbu and N. Limnios. Semi-Markov Chains and Hidden Semi-Markov Models Toward Applications. Springer-Verlag New York Inc, 2008. 

\bibitem{dami12c} G. D'Amico, F. Petroni, A semi-Markov model for price returns. {\em Physica A} $\bf 391$ (2012) 4867-4876.

\bibitem{ciardo90} G. Ciardo, R. Marie, B. Sericola, K.S. Trivedi, Performability Analysis using Semi-Markov Reward Processes, {\em IEEE Transactions on Computers}, $\bf C-39$(10) (1990) 1251-1264.

\bibitem{dami09b} G. D'Amico, The crossing barrier of a non-homogeneous semi-Markov chain, {\em Stochastics: An International Journal of Probability and Stochastic Processes}, $\bf 81$(6) (2009) 589-600.

\bibitem{csenk95} A. Csenki, Transition analysis of semi-Markov reliability models - a tutorial review with emphasis on discrete-parameter approaches, In S. Osaki, editor, {\em Stochastic Models in Reliability and Maintenance,} (2002) 219-251, Springer, Berlin.

\bibitem{dami11c} 
G. D'Amico, J. Janssen, and R. Manca, Duration Dependent Semi-Markov Models, {\em Applied Mathematical Sciences} $\bf 5$(42) (2011) 2097-2108.

\bibitem{dami13c} G. D'Amico, F. Petroni and F. Prattico, First and second order semi-Markov chains for wind speed modeling, {\em Physica A} $\bf 392$ (2013) 1194-1201.

\bibitem{barl75} R.E. Barlow, F. Prochan, {\em Statistical Theory of Reliability and Life Testing: Probabilistic Models.} Holt, Rinehart and Winston, New York, 1975. 

\bibitem{lisn03} G. Lisnianski, G. Levitin, {\em Multi-State System Reliability.} World Scientific, N.J. (2003). 

\bibitem{pere02} R. Perez-Ocon, I. Torres-Castro, A reliability semi-Markov Model involving geometric processes, {\em Applied Stochastic Models in Business and Industry} $\bf 18$(2) (2002) 157-170.

\bibitem{limn99}
N. Limnios, G. Opri\c{s}an,  A unified approach for reliability and performability evaluation of semi-Markov systems, {\em Applied Stochastic Models in Business and Industry} $\bf 15$ (1999) 353-368.




\end{thebibliography}
\end{document}